\documentclass[journal,magazines]{IEEEtran}

\usepackage{graphicx}
\usepackage{booktabs}
\usepackage{amsmath}
\usepackage{amssymb}
\usepackage[colorlinks]{hyperref}

\title{Fluid RIS (FRIS)-Assisted Index Modulation \\ for 6G Wireless Communications}

\author{Xusheng Zhu,
Kai-Kit Wong,
Sai Xu,
Hao Xu,
Wen Chen, and
Hyundong Shin\thanks{X. Zhu, K. K. Wong, S. Xu are with the Department of Electronic and Electrical Engineering, University College London, London WC1E 7JE, United Kingdom (e-mail: xusheng.zhu@ucl.ac.uk; kai-kit.wong@ucl.ac.uk). K. K. Wong is also affiliated with the Department of Electronic Engineering, Kyung Hee University, Yongin-si, Gyeonggi-do 17104, South Korea. H. Xu is with the National Mobile Communications Research Laboratory, Southeast University, Nanjing 210096, China (e-mail: hao.xu@seu.edu.cn). W. Chen is with the Department of Electronic Engineering, Shanghai Jiao Tong University, Shanghai 200240, China (e-mail: wenchen@sjtu.edu.cn). Hyundong Shin is with the Department of Electronics and Information Convergence Engineering, Kyung Hee University, Yongin-si, Gyeonggi-do 17104, Republic of Korea (e-mail: hshin@khu.ac.kr).}

\thanks{\em Corresponding author: Kai-Kit Wong.}
}

\begin{document}
\maketitle
\begin{abstract}
Fluid reconfigurable intelligent surfaces (FRIS) extend conventional reconfigurable intelligent surfaces (RIS) by adding spatial reconfigurability through switchable apertures, pattern-reconfigurable units, fluidic conductive materials, or movable surface elements. This article studies how FRIS can support index modulation (IM), where information bits select a surface configuration and the receiver detects the index from the induced receiver-side response. A key challenge is that many feasible FRIS layouts do not necessarily lead to many reliable spatial indices. After propagation, mutual coupling, hardware distortion, and receiver observation, different layouts may produce similar receiver-side responses and cause index-detection errors. To address this issue, we present a response-aware design view, in which FRIS spatial codebooks are selected according to response-domain separability rather than layout diversity alone.
We also discuss actuation granularity as a practical design knob that balances spatial diversity, pilot overhead, coupling robustness, and hardware feasibility. The resulting workflow helps select compact, trainable, and controllable spatial-index codebooks from dense FRIS layouts, providing design guidance for future programmable wireless environments.
\end{abstract}

\section{Introduction}
\IEEEPARstart{F}{uture} sixth-generation (6G) wireless networks are expected to support reliable, high data-rate, and low-latency links in dynamic and blockage-prone environments~\cite{Wang_COMST_2023}. This is especially challenging at millimeter-wave and terahertz bands, where channels are sensitive to path loss, mobility, blockage, and device orientation. Reconfigurable intelligent surfaces (RIS) can help by making the wireless environment more controllable through programmable electromagnetic responses~\cite{Pan_Mag_2021}. However, conventional RIS usually has a fixed geometry. Its phase or amplitude response can be tuned, but its spatial layout remains unchanged.

At the same time, index modulation (IM) shows that information can also be conveyed by the indices of available transmission resources, rather than only by amplitude, phase, or frequency~\cite{Basar_Access_2017}. This naturally leads to the following question:
\vspace{1mm}
\begin{quote}
{\em Can a programmable wireless environment also provide useful indices for information transfer?}
\end{quote}
\vspace{1mm}

\begin{table*}[!t]
\centering
\caption{From RIS to FRIS-Assisted IM}
\label{tab:ris_fris_im}
\renewcommand{\arraystretch}{1.02}
\setlength{\tabcolsep}{7pt}
\footnotesize
\begin{tabular}{lccc}
\toprule
\textbf{Aspect} & \textbf{RIS} & \textbf{FRIS} & \textbf{FRIS-Assisted IM} \\
\midrule
Main role & Reflection control & Channel adaptation & Information mapping \\
Controlled resource & Phase/amplitude & Phase/amplitude + layout & Configuration index \\
Information carrier & Waveform & Waveform/channel & Spatial index \\
Key design issue & Channel gain & Channel matching & Response separability \\
Main constraint & Fixed geometry & Coupling/control cost & Codebook reliability \\
\bottomrule
\end{tabular}
\end{table*}

\begin{figure*}[]
\centering
\includegraphics[width=0.86\textwidth]{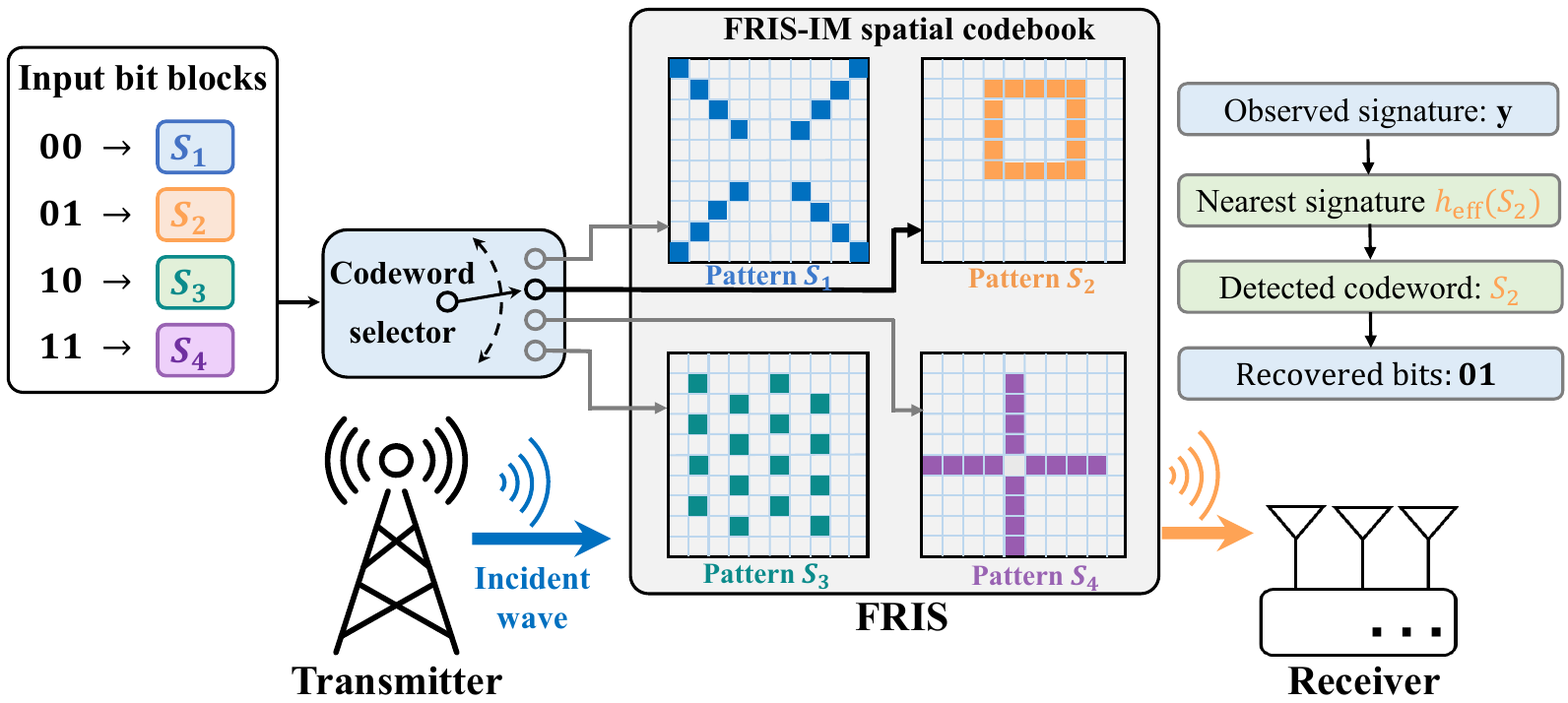}
\caption{Basic principle of FRIS-assisted IM.
Input bits select one FRIS spatial codeword, which configures the active or effective surface regions over a reconfiguration interval.
The receiver detects the spatial index by matching the observed signature with the expected codeword-dependent responses.}
\label{fig:fris_im_principle}
\end{figure*}
The fluid antenna system (FAS) concept shows that position reconfigurability can create new spatial opportunities beyond fixed layouts~\cite{Wong_TWC_2021,New-2025comst,Zhu_arXiv_2025_FAS}. Inspired by this idea, fluid reconfigurable intelligent surfaces (FRIS) bring position or pattern reconfigurability into RIS architectures~\cite{Ghadi_WCL_2025_FIRES,Xiao_TWC_2026,Zhu_WCL_2026,Xiao_arXiv_2025_FRIS}. In this article, we use FRIS in a broad sense. Its spatial state may be realized by switchable effective apertures, pattern-reconfigurable units, fluidic conductive materials, or movable surface elements. The key feature is that FRIS can change not only its electromagnetic response, but also the spatial layout or effective active regions seen by the incident wave.

This spatial freedom enables FRIS-assisted IM. Information bits select one configuration from a predefined FRIS spatial codebook. The selected configuration produces a receiver-side response, and the receiver detects the transmitted index by matching the observed response with the expected codeword-dependent responses, as shown in Fig.~\ref{fig:fris_im_principle}. The FRIS index does not replace conventional waveform modulation. Instead, it adds a spatial-index layer over a reconfiguration interval, while waveform symbols can still carry data within the same interval. This idea is useful in blockage-prone indoor millimeter-wave or terahertz links, where FRIS configurations may create distinguishable responses when direct paths are weak or intermittent. It is also useful for low-power or battery-constrained IoT systems, where a slowly updated surface-state index can complement conventional modulation without adding active RF chains at the device side.

The main challenge is that many feasible FRIS layouts do not imply many reliable spatial indices. Two layouts may look different on the surface, but after propagation, mutual coupling, hardware distortion, and receiver observation, they may produce similar responses at the receiver. The receiver may then confuse their indices. Dense placement can also increase near-field mutual coupling, while a large configuration space increases pilot training, control signaling, and reconfiguration overhead~\cite{Gradoni_LWC_2021,Qian_LWC_2021,An_MWC_2024}. Therefore, practical FRIS-assisted IM should not simply use as many physical layouts as possible. It should select configurations whose receiver-side responses are easy to distinguish, train, and control.

Table~\ref{tab:ris_fris_im} shows how the role of the surface changes from RIS to FRIS-assisted IM. RIS mainly controls reflection over a fixed geometry. FRIS adds spatial reconfigurability for channel adaptation. FRIS-assisted IM further uses the selected surface configuration as an information-bearing index. Unlike fixed-geometry RIS-based IM, where the index is usually associated with phase states, sub-surface activation, or reflection patterns, FRIS-assisted IM introduces geometry- or pattern-dependent indices~\cite{Zhu_JSAC_2025,Zhang_arXiv_2026}. Their reliability depends not only on reflection coefficients, but also on aperture reshaping, mutual coupling, calibration accuracy, actuation granularity, and reconfiguration feasibility.

This article develops a response-aware design view for practical FRIS-assisted IM. The main message is simple: FRIS spatial freedom is useful for IM only when it creates receiver responses that can be reliably separated. First, we explain why the FRIS spatial codebook should be designed according to receiver responses, rather than layout diversity alone. Second, we discuss actuation granularity as a practical design knob that links spatial diversity, mutual coupling, pilot overhead, and hardware feasibility. Third, we present a workflow that selects compact, response-separable, trainable, and controllable spatial-index codebooks from feasible FRIS configurations.


\section{FRIS-Assisted IM: From Surface Layouts to Spatial Indices}

FRIS-assisted IM maps information bits to selected surface configurations. The controller stores a finite set of feasible FRIS configurations as a spatial codebook. During one reconfiguration interval, one codeword \(\mathcal{S}_i\) is selected. This codeword determines which regions of the FRIS are activated, switched, or moved, and it produces a corresponding response at the receiver. The receiver detects the spatial index by comparing the observed signal with the expected responses in the codebook.

\subsection{Spatial-Indexing Mechanism}

For a selected FRIS codeword \(\mathcal{S}_i\), the received signal over the corresponding reconfiguration interval can be written as
\[
    \mathbf{y}
    =
    \mathbf{h}_{\mathrm{eff}}(\mathcal{S}_i)x
    +
    \mathbf{n},
\]
where \(x\) is the waveform-domain signal, \(\mathbf{n}\) is noise, and \(\mathbf{h}_{\mathrm{eff}}(\mathcal{S}_i)\) is the receiver-side response induced by \(\mathcal{S}_i\). This response may be a scalar channel coefficient, a multi-antenna response vector, or a feature extracted from received samples. If the codebook contains \(K\) codewords that can be reliably distinguished, it can carry \(\log_2 K\) spatial-index bits per reconfiguration interval. Therefore, the useful index gain depends not only on the codebook size, but also on the interval duration, pilot overhead, and hardware cost needed to maintain the codebook.

\subsection{Layout Domain versus Response Domain}

A FRIS layout describes the physical state of the surface. It specifies which elements, groups, or effective regions are activated, switched, or moved. The receiver response describes what the receiver actually observes for that layout. These two domains are related, but they are not the same.

Two layouts may look different on the surface, but they may still produce similar responses at the receiver. This can happen because of propagation, mutual coupling, calibration errors, and receiver observation. Such layouts are unreliable codeword candidates, since the receiver may confuse their indices. For this reason, a FRIS-assisted IM codebook should not be selected only according to layout diversity. It should also consider whether the corresponding receiver responses are easy to distinguish.

\subsection{Response Separability}

To detect the spatial index reliably, different codewords should produce well-separated receiver responses. For two FRIS configurations \(\mathcal{S}_i\) and \(\mathcal{S}_j\), the response-domain distance can be measured as
\[
d_{i,j}
=
\left\|
\mathbf{h}_{\mathrm{eff}}(\mathcal{S}_i)
-
\mathbf{h}_{\mathrm{eff}}(\mathcal{S}_j)
\right\|^2.
\]
A small \(d_{i,j}\) means that the two indices are difficult to distinguish, even when their physical layouts are different. A large \(d_{i,j}\) means that the receiver can separate them more reliably. We refer to this property as response-domain separability, or simply response separability.

A response-aware FRIS-IM codebook can then be constructed by selecting a subset of feasible configurations that maximizes the minimum pairwise response distance:
\[
\mathcal C^\star =
\arg\max_{\mathcal C\subseteq\mathcal S_{\rm cand},|\mathcal C|=K}
\min_{\substack{\mathcal{S}_i,\mathcal{S}_j\in\mathcal C\\ i\neq j}}
d_{i,j},
\]
where \(\mathcal S_{\rm cand}\) is the feasible configuration set and \(K\) is the target codebook size. This formulation does not require all feasible layouts to be used as indices. Instead, it keeps the configurations that generate well-separated receiver responses.

\begin{figure*}[!t]
    \centering
    \includegraphics[width=0.98\textwidth]{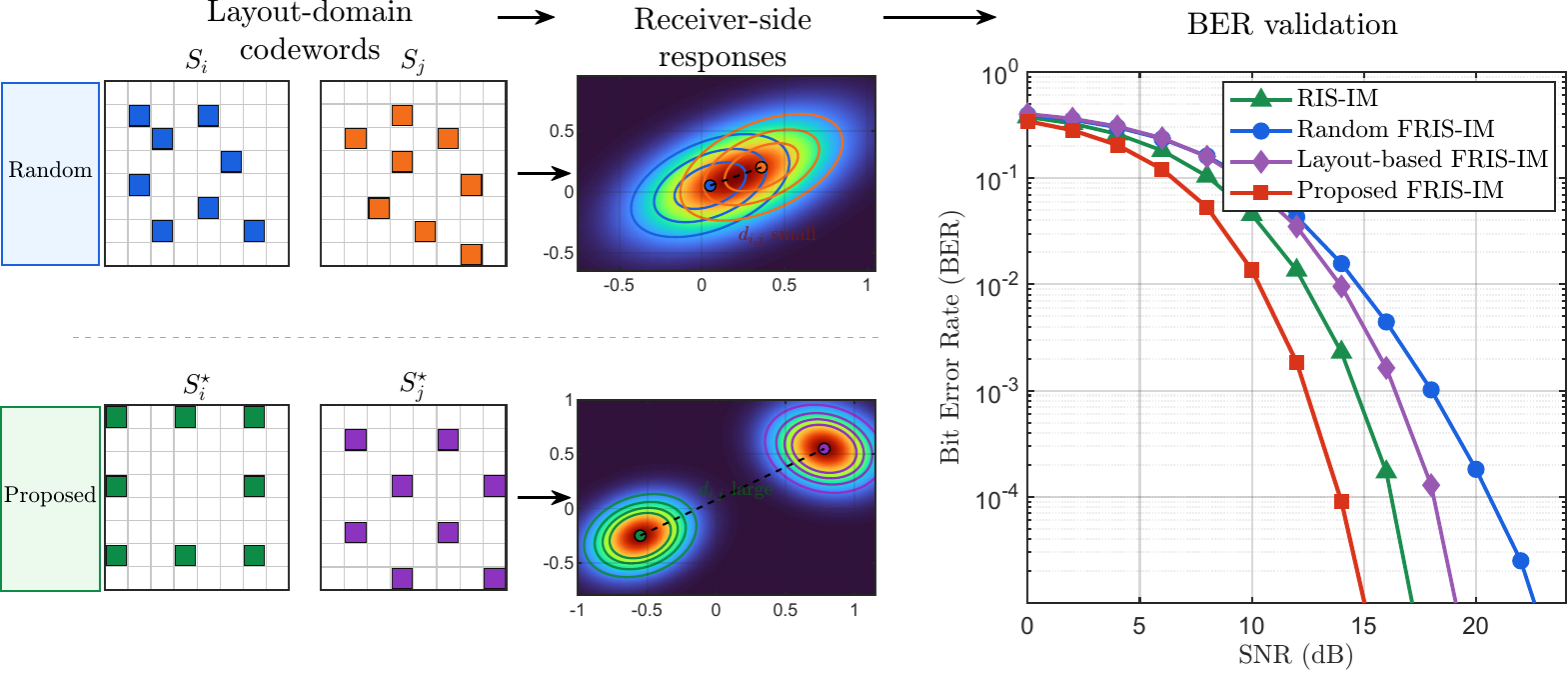}
\caption{Response-aware FRIS-IM codebook design: from layout-domain selection to BER validation.
Random FRIS-IM may select visually distinct layouts that generate overlapping receiver-side response signatures after propagation and coupling, while layout-based FRIS-IM uses physical layout separation as a baseline.
The proposed FRIS-IM enlarges the minimum response-domain distance and improves BER under equal codebook size and equal active-unit cardinality.}
\label{fig:response_aware_codebook_ber}
\end{figure*}

Fig.~\ref{fig:response_aware_codebook_ber} illustrates this idea. A random FRIS-IM codebook may select visually distinct layouts without checking their receiver responses. As a result, different codewords may still overlap in the response domain. A layout-based FRIS-IM codebook improves the selection by considering physical layout separation. However, layout diversity alone cannot guarantee response separability after propagation, coupling, and receiver observation. In contrast, the proposed response-aware FRIS-IM codebook selects feasible configurations according to their receiver-side response distances, which leads to more separated responses and more reliable index detection.

The BER curves in Fig.~\ref{fig:response_aware_codebook_ber} are illustrative and are used to highlight the role of response separability in index detection. The RIS-IM curve represents a fixed-geometry RIS-based IM benchmark with no layout reconfiguration. Random FRIS-IM randomly selects feasible FRIS configurations, while layout-based FRIS-IM selects configurations according to physical layout separation. The proposed FRIS-IM uses receiver-side response separability for codebook construction. Under the same codebook size and the same active-unit cardinality, the proposed response-aware codebook achieves a lower BER because it increases the minimum response distance between spatial-index codewords.

\subsection{Practical Codebook Constraints}

The above discussion assumes that the receiver responses can be estimated or calibrated. In practice, the selected codebook must also satisfy hardware and protocol constraints. These constraints include actuation range, switching speed, energy budget, calibration accuracy, hardware response time, CSI update period, channel coherence time, pilot budget, and service latency.

Fast switching may allow more configurations to be trained and updated within a short time. Fluidic or mechanical movement may instead favor fewer and more stable states. Therefore, the goal is not to use all layouts generated by a dense aperture. The goal is to select configurations whose receiver responses remain separable, trainable, and controllable under electromagnetic, pilot, control, and hardware constraints.

\section{Near-Field Mutual Coupling in Dense FRIS}

The response-aware design in Fig.~\ref{fig:response_aware_codebook_ber} raises a basic physical question: why can two different FRIS layouts produce similar receiver responses? In dense FRIS implementations, one important reason is near-field mutual coupling. When elements or effective reflecting regions are placed close to each other, the current and field induced on one unit can affect its neighbors. As a result, the surface behaves as a coupled electromagnetic structure, rather than a set of independent controllable points.

\subsection{Coupling-Induced Ambiguity}

For FRIS-assisted IM, mutual coupling changes the mapping from a selected surface configuration to the received response. Energy may leak from selected regions to nearby regions, and the field generated by one active part may be distorted by adjacent units. Therefore, the effective response \(\mathbf{h}_{\mathrm{eff}}(\mathcal{S}_i)\) is determined by the coupled behavior of the selected regions, not by each region independently.

This effect can reduce the response distance \(d_{i,j}\) between two codewords. When \(d_{i,j}\) becomes small, the receiver may confuse the corresponding spatial indices, even if the two layouts look different on the surface. For this reason, mutual coupling should be considered during codebook design. It should not be treated only as an impairment to be compensated after detection.

\subsection{Intra-Group and Inter-Group Coupling}

A practical way to handle coupling is to separate it into intra-group and inter-group effects. Intra-group coupling occurs inside the same controlled unit. If the local structure is fixed, or if it belongs to a small set of calibrated states, this effect can be included in an equivalent response obtained by simulation, measurement, or offline calibration. After this calibration, the group can be treated as one controllable unit in codebook design.

Inter-group coupling occurs between different selected groups or controlled units. It is more harmful for index detection, because one selected unit can change the response of another. If two selected groups are too close, their combined response may become unstable or may look similar to that of another codeword. This reduces the number of useful spatial indices and motivates spacing-aware as well as response-aware codebook selection.

\subsection{Geometric Isolation and Codebook Selection}

Geometric isolation provides a simple first-order way to reduce harmful inter-group coupling. The FRIS codebook should avoid selecting strongly interacting groups at the same time. A center-to-center spacing on the order of half a wavelength can be useful in many designs, although the exact value depends on the element geometry, polarization, substrate, packaging, carrier frequency, and propagation environment.

This rule does not remove all coupling. Instead, it makes the codebook design more manageable. Dense FRIS positions should therefore be treated as candidate locations, not as independent spatial indices that must all be used. The final codebook should be selected according to both physical feasibility and receiver-side response separability.

\begin{figure*}[!t]
    \centering
    \includegraphics[width=0.82\textwidth]{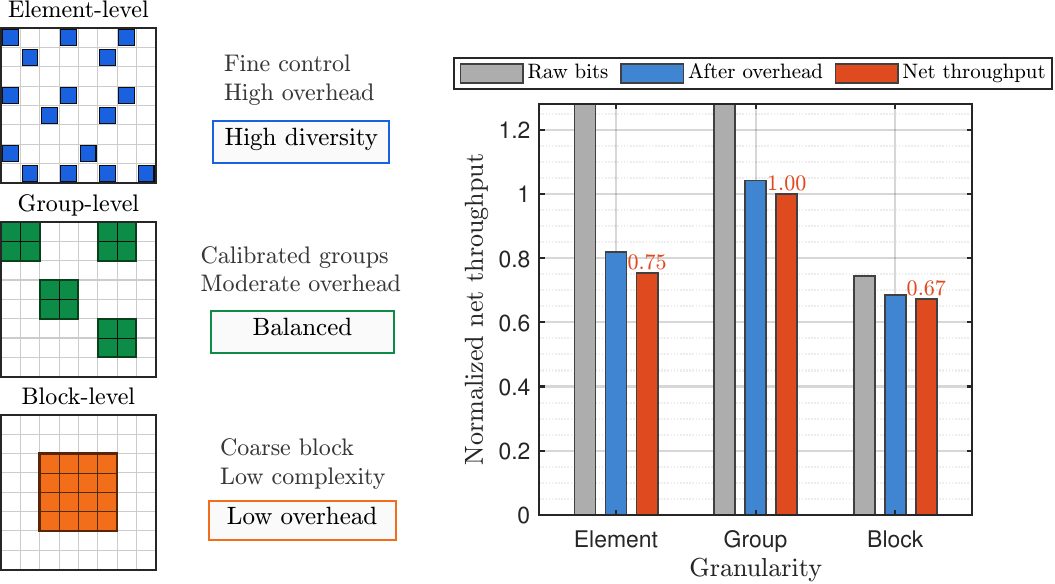}
    \caption{Control granularity and net-throughput tradeoff for FRIS-assisted IM.
All three modes use the same aperture and activated-element cardinality, \(N_{\rm act}=16\).
The bar chart compares raw index bits, overhead-penalized bits, and net throughput.
Group-level control provides the best overhead-aware tradeoff in this illustrative setting.}
\label{fig:granularity_tradeoff}
\end{figure*}
\section{Managing Spatial Degrees of Freedom (DoFs) Through Actuation Granularity}

The previous section shows that dense FRIS positions should not be treated as independent spatial indices without considering coupling and response separability. Another important factor is how the surface is controlled. Different FRIS implementations may support different control units and update speeds. Electronic or pixel-level switching may allow fine and fast pattern changes, while fluidic or mechanical movement may favor coarser and more stable configurations.

This motivates the concept of actuation granularity, which describes the basic level at which the FRIS is activated, switched, moved, or reconfigured. In FRIS-assisted IM, this granularity is not only a hardware choice. It also affects the candidate layout space, pilot overhead, coupling behavior, update rate, and index-detection complexity.

\subsection{Element-, Group-, and Block-Level Control}

Fig.~\ref{fig:granularity_tradeoff} compares three representative choices: element-level, group-level, and block-level control. All three modes use the same aperture and the same activated-element cardinality. They should be viewed as scenario-dependent design options, rather than as a fixed performance ranking.

Element-level control treats each element or pixel as an independent controllable unit. It provides fine spatial resolution and a large candidate layout space. It is useful when fast switching, accurate calibration, and sufficient pilot resources are available. Its main cost is high training overhead, stronger coupling uncertainty, and higher detection complexity.

Group-level control activates calibrated groups of elements or candidate positions. The response inside a group can be measured, simulated, or calibrated as an equivalent response. At the same time, selected groups can be spaced apart to reduce harmful inter-group coupling. This option is useful when the system needs to balance response diversity, pilot overhead, coupling robustness, and hardware feasibility.

Block-level control uses a larger active region as one controllable unit. It reduces training, calibration, and actuation cost, and is useful for low-complexity systems or slow-reconfiguration hardware. Its main limitation is a smaller effective codebook, because fewer distinguishable receiver responses can be generated.

\subsection{Granularity-Dependent Net-Throughput Tradeoff}

The choice of actuation granularity affects both communication performance and implementation cost. Small control units provide more candidate configurations, but they require more pilots, more accurate coupling calibration, and more complex control. Large control units reduce overhead and are easier to actuate, but they may also reduce the number of useful spatial indices. Thus, the useful spatial DoFs are not simply the number of physical positions on the aperture. They are the configurations that remain separable, trainable, and feasible under the given system constraints.

To make this tradeoff clear, the right-hand side of Fig.~\ref{fig:granularity_tradeoff} reports three normalized quantities for each actuation granularity. The gray bar, ``Raw bits'', denotes the nominal spatial-index payload \(\log_2(K_{\rm eff})\), where \(K_{\rm eff}\) is the effective response-separable codebook size. The blue bar, ``After overhead'', accounts for pilot training, control signaling, and reconfiguration overhead through the factor \(1-T_{\rm oh}/T_c\). The red bar, ``Net throughput'', further includes the effect of index-detection reliability through the factor \(1-P_e\). The normalized net spatial-index throughput is written as
\[
\bar R_{\rm net}
=
\left(1-\frac{T_{\rm oh}}{T_c}\right)
\log_2(K_{\rm eff})(1-P_e),
\]
where \(T_{\rm oh}/T_c\) is the normalized overhead ratio and \(P_e\) is the spatial-index detection error probability. This metric is used to illustrate the overhead-aware spatial-index tradeoff, not to provide a complete capacity analysis.

The result in Fig.~\ref{fig:granularity_tradeoff} shows that the largest raw index space does not necessarily give the largest practical throughput. Element-level control has the largest nominal index payload, but its high overhead reduces the net throughput. Block-level control has low overhead, but its smaller effective codebook limits the index rate. Group-level control provides an intermediate design. It keeps enough response diversity through calibrated groups while reducing pilot and control overhead. In this illustrative setting, it provides the best overhead-aware tradeoff.
\begin{figure*}[!t]
    \centering
    \includegraphics[width=0.9\textwidth]{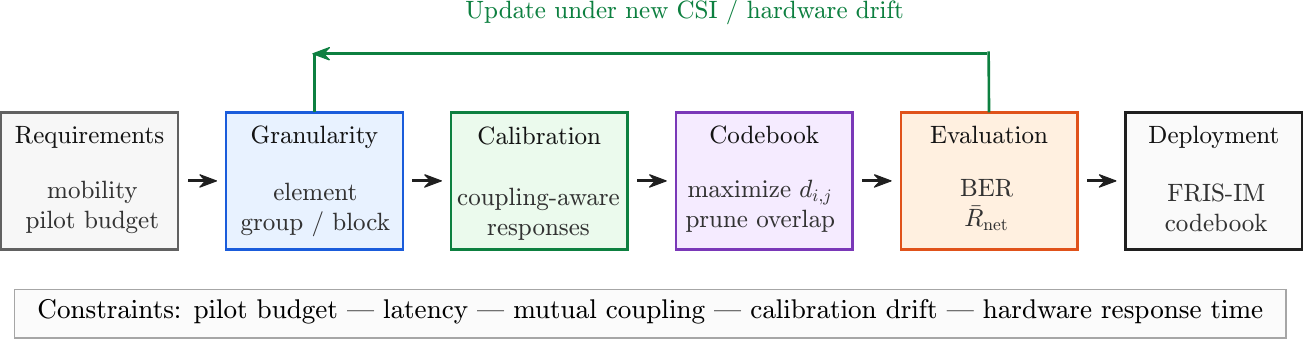}
    \caption{Practical response-aware design workflow for FRIS-assisted IM.
The workflow starts from system requirements, selects actuation granularity, calibrates coupling-aware responses, constructs a response-separable codebook, evaluates BER and net throughput, and deploys a compact FRIS-IM codebook.
The feedback loop updates the design under new CSI, hardware drift, and practical constraints.}
\label{fig:workflow}
\end{figure*}
\subsection{Design Bridge}

Actuation granularity links the FRIS hardware to the communication design. It determines the candidate layout space, pilot and control overhead, coupling behavior, update rate, and detection complexity. Fine granularity can increase nominal spatial freedom, but it may be hard to train and control. Coarse granularity can improve robustness and reduce latency, but it may limit the index rate. Therefore, granularity should be selected together with response-aware codebook construction. A deployable FRIS-assisted IM system should first determine the useful response-separable spatial DoFs under pilot, latency, coupling, and hardware constraints, and then choose the actuation granularity and codebook size accordingly.

\section{Practical Response-Aware Workflow for FRIS-Assisted IM}

The preceding sections show that practical FRIS-assisted IM should not start by enumerating all possible surface layouts. It should start from the receiver responses that can be reliably separated, trained, and controlled under coupling, pilot, latency, and hardware constraints. Fig.~\ref{fig:workflow} summarizes a compact workflow that turns system requirements into a deployable FRIS-IM spatial codebook.

\textit{Requirements:} The design first specifies the main operating conditions, including mobility, coherence time, pilot budget, latency target, and hardware response time. These factors determine how often the FRIS state can be updated, how many configurations can be trained, and how large the spatial codebook can be.

\textit{Granularity selection:} The system then chooses element-, group-, or block-level actuation according to the pilot budget, switching speed, coupling robustness, and implementation complexity. As discussed in Fig.~\ref{fig:granularity_tradeoff}, the preferred granularity depends on the scenario and should not be treated as universal.

\textit{Response calibration:} After the granularity is chosen, the receiver responses of the controllable units should be calibrated. This calibration should account for propagation, mutual coupling, receiver observation, and hardware-dependent effects. The resulting response map can be obtained through electromagnetic simulation, measurement, or offline training.

\textit{Codebook construction:} The codebook is then built from feasible configurations that satisfy movement limits, spacing rules, actuation energy, and reconfiguration latency. Among these candidates, response-aware selection keeps codewords with large pairwise response distances and removes configurations that produce overlapping receiver responses.

\textit{Evaluation and update:} The selected codebook is evaluated using index-detection reliability, pilot overhead, latency, energy consumption, and net throughput. When CSI, hardware states, or service requirements change, the system may reduce the codebook size, adjust spacing or grouping, or switch to another actuation granularity.

The workflow follows a simple response-to-layout rule. The designer should first identify receiver responses that can be separated, trained, and controlled, and then map these useful responses back to feasible FRIS layouts. This avoids two extremes: using every dense-grid position as an index, which creates high training and coupling cost, and using overly coarse control, which loses most of the spatial-index diversity.

\section{Future Vision and Open Challenges}

Based on the response-aware workflow above, several issues remain before FRIS-assisted IM can be deployed in practical systems. The surface is no longer used only to shape the channel; it can also provide spatial indices for information transfer. To make this practical, the system must keep the receiver responses separable while controlling pilot overhead, reconfiguration latency, hardware uncertainty, and implementation cost. This leads to several open challenges in codebook learning, channel tracking, scheduling, hardware design, and multi-user or multi-functional operation.

\subsection{Low-Overhead Response-Aware Codebook Learning}

A practical FRIS-assisted IM system needs a compact codebook with well-separated receiver responses. However, testing all feasible configurations is not scalable for dense apertures. Learning-based methods can reduce the search space by using partial CSI, user location, past measurements, and hardware states. Different from conventional RIS optimization, which mainly aims to improve channel gain, FRIS-assisted IM learning should also preserve pairwise response separability. The learning process should therefore account for coupling, spacing constraints, movement limits, and hardware response time.

\subsection{Channel Tracking and Codeword Calibration}

The mapping from a FRIS configuration to its receiver response may change over time because of mobility, blockage, platform motion, temperature drift, actuator errors, or hardware aging. In static deployments, offline calibration and occasional pilots may be sufficient. In mobile or blockage-prone links, lightweight tracking is needed. A promising direction is to track only the useful response features rather than retrain the full configuration space. This can be done through hierarchical training or predictive update of a small candidate codebook.

\subsection{Scheduling Under Reconfiguration Latency}

The useful index rate depends on how fast the FRIS state can be trained, selected, and updated. Fast electronic or pixel-level switching may support short reconfiguration intervals, while fluidic or mechanical movement may require longer intervals. Therefore, FRIS-assisted IM should be viewed as a slower spatial-index layer that complements conventional waveform modulation within each reconfiguration interval. Future schedulers should jointly choose the update interval, active codebook size, and number of spatial indices according to coherence time, traffic load, latency target, and hardware response time.

\subsection{Robust Hardware and Reliable Spatial-Index Realization}

A practical FRIS must realize each selected codeword accurately and repeatedly. Movement errors, actuator delay, element misalignment, packaging limits, substrate loss, and environmental changes may perturb the receiver response. These errors can shrink the response margin between codewords and increase index-detection errors. For this reason, codebook construction should consider not only nominal response separability, but also robustness to hardware uncertainty, movement cost, energy consumption, and long-term drift. An important open problem is how to translate hardware error limits into minimum response-distance requirements.

\subsection{Multi-User and Multi-Functional FRIS-Assisted IM}

Future FRIS systems may support multiple users, simultaneous reflection and transmission, or integrated sensing and communication (ISAC). In these settings, one surface state can create different responses for different users or tasks. A codeword that is easy to detect for one receiver may be ambiguous for another. For ISAC-oriented deployments, FRIS codewords may need to generate distinguishable communication-channel responses while also forming sensing beams with low ambiguity or controlled target illumination. This leads to a joint codebook and scheduling problem that balances response separability, sensing-beam orthogonality, multi-user interference, fairness, and energy cost.

\section{Conclusion}

FRIS-assisted IM provides a new way to encode information through programmable wireless environments by mapping bits to selected surface configurations.
The main value of this approach is not to generate as many physical layouts as possible, but to create receiver responses that can be reliably separated, trained, and controlled under coupling, pilot, latency, and hardware constraints.
This article presented a response-aware design view, where FRIS spatial codebooks are selected according to response separability rather than layout enumeration alone.
We also showed that actuation granularity is a practical design knob that links spatial diversity, mutual coupling, pilot overhead, and hardware feasibility.
Element-, group-, and block-level control should therefore be treated as scenario-dependent choices rather than a fixed ranking.
The main lesson is that future programmable environments should not only shape wireless channels, but also create reliable spatial indices for information transmission.

\end{document}